\documentclass[prb,aps,notitlepage, twocolumn,superscriptaddress]{revtex4-1}
\usepackage{amsmath,amssymb}
\usepackage{graphicx}
\usepackage{epsfig, color}
\usepackage{wasysym}
\usepackage{amsfonts}
\usepackage{bm}
\usepackage{enumerate}
\usepackage{color}
\usepackage[normalem]{ulem}
\usepackage{cancel}

\usepackage{latexsym}
\usepackage[breaklinks,colorlinks = true,linkcolor = red,urlcolor=cyan,citecolor=red]{hyperref}

\newcommand{\be}{\begin{eqnarray}}
\newcommand{\ee}{\end{eqnarray}}
\newcommand{\bent}{\begin{equation*}}
\newcommand{\eent}{\end{equation*}}
\newcommand{\bw}{\begin{widetext}}
\newcommand{\ew}{\end{widetext}}

\newcommand{\expval}[1]{ \left\langle #1 \right\rangle}

\newcommand{\abs}[1]{ \left| #1 \right|}

\newcommand{\vpd}{\vphantom{\dagger}}
\newcommand{\vps}{\vphantom{s}}

\def\dof{d.o.f.}
\def\coup{\tilde{\eta}}

\begin{document}
\title{Dissipative Luttinger liquids}
\author{Aaron J. Friedman}
\affiliation{Rudolf Peierls Centre for Theoretical Physics, University of Oxford, OX1 3NP, UK}
\affiliation{Department of Physics, University of Massachusetts, Amherst, Massachusetts 01003, USA}
\affiliation{Department of Physics and Astronomy, University of California, Irvine, CA 92697, USA}

\date{\today}

\begin{abstract}
We investigate a one dimensional quantum fluid coupled to a dissipative bath. The quantum fluid is captured by the canonical Luttinger liquid; the bath is given by the model of Caldeira and Leggett, i.e. a tower of oscillators coupled linearly to the fluid density, $\rho$. The bath can be integrated out exactly, producing an effective interaction for the fluid that is nonlocal in time; we argue that the form corresponding to Ohmic dissipation is generic. Compared to previous works, we compute correlation functions for this minimal model without approximation, including at finite temperature $T>0$. From these and a Kubo calculation, we conclude that \emph{arbitrary} dissipation destroys the perfect conductivity of the Luttinger liquid via Zeno localization, even in the absence of a spatial potential; from RG analysis of harmonic terms, we also find that the open Luttinger liquid is significantly more prone to localization by such potentials, in contrast to the usual intuition that baths make systems \emph{less} localized.
\end{abstract}

\maketitle

\section{Introduction} \label{sec:Intro}
The ability to realize quantum systems that are well isolated from their environment has led to new efforts to understand how such systems thermalize. 
The Eigenstate Thermalization Hypothesis (ETH)\cite{ETH1,ETH2}~provides a mechanism for part of an isolated system to equilibrate with the rest, which acts as a bath (thermal reservoir), to reproduce the familiar results of statistical mechanics. 
These efforts have also revealed classes of non-ergodic systems that do not thermalize per ETH.

In practice, no experiment can remain isolated forever\cite{RevModPhys.80.885,PhysRevLett.120.160404,MBLopenExp}, which motivates careful theoretical study of the environment itself\cite{MBLBath,GenMBLBath,2014arXiv1402.5971N,hyattSmallBath}. Such studies have focused primarily on  Markovian (i.e. memoryless) baths, whose action is history-independent; such baths generally delocalize systems, rendering them more thermal\cite{GenMBLBath}. 

A fundamental result of ETH is that the mechanism for thermalization of a [sub-]system is \emph{entanglement} between the `system' and `bath' \cite{ETH1,ETH2}. Thus, 
a Markovian bath---which foregoes any description of the bath itself---necessarily overlooks this key aspect of thermalization. 
In fact, the failure of quantum systems to thermalize is highly dependent on such details; hence, Markovian baths may not be ideal for understanding possible dynamics beyond outright thermalization.

To gain insight into alternative behavior in the presence of baths, we turn to techniques used to study quantum dissipation; in particular, the model pioneered by Caldeira and Leggett (CL)\cite{CL1,CL2} and used in numerous subsequent works\cite{Guinea1,FZ,CL4,LeggettTwoState,OLL,QBM}. 
Although CL baths are designed to thermalize a system (c.f. a standard Keldysh calculation\cite{KamenevKeldyshRef}), recent work has shown that it is nonetheless possible for system coupled to these baths to feature dynamical properties reminiscent of localization\cite{FZ,QBM,OLL}. 
A single particle, moving in a harmonic potential landscape in the presence of such dissipation will undergo a dynamical localization transition as the dissipation strength is \emph{increased}\cite{FZ}, and the decohering effect of the bath results in effective Zeno localization\cite{Anderson2Zeno} to one of the potential wells; recently, it was shown that the inclusion of a second, incommensurate potential will destroy the delocalized phase entirely, for arbitrary non-zero coupling to the bath\cite{QBM}. 

Most studies of CL baths involve non-interacting systems, or a single particle; despite these simplified settings, such systems can nonetheless exhibit phase transitions\cite{LeggettTwoState,Guinea1,FZ,CallanFreed1,CallanFreed2,WEISS198563,KaneYiQBM,ASLANGUL1,PhysRevLett.94.047201}. 
However, exact results for interacting systems have remained largely elusive.

In this work, we consider the effect of such a dissipative bath on an interacting, one-dimensional quantum system, which in various limits may represent any of a number of experiments. 
Taking the system to be somewhat insulated from its environment, one expects the bath to have a weak, dissipative effect thereupon, motivating the use of the CL formalism. 
Since the CL bath equilibrates with the system, we exploit standard equilibrium methods to examine the low-temperature properties of the combined system and bath, and also obtain results for arbitrary $T>0$ using the Matsubara formalism.

At long wavelengths, the physics of fluids comprising either boson or fermion degrees of freedom (\dof)~is captured by the paradigmatic Luttinger liquid \cite{Luttinger,MattisLieb,LutherPeschel,Haldane2,HaldaneLuttinger,giamarchi_book,vDS}. 
The bare action \eqref{eq:LLactionxt} for this system is quadratic in a scalar, bosonic `displacement' field, $\phi \left( x, \tau \right)$, related to fluctuations of the density. 
Depending on microscopic details, there may also be corrections of the form $\cos \left[ m \phi \left( x, \tau \right) \right]$, for various harmonics $m$, which may correspond to spatial potentials, Umklapp processes, etc. 
We comment on the physical relevance of these terms using correlation functions and a renormalization group (RG) analysis.

Regarding the partition function of the Luttinger liquid and bath, we can trace out the bath \dof~exactly\cite{FeynmanVernon,CL1,CL2,CL4,LeggettTwoState,FZ,QBM,Guinea1,OLL} to obtain an effective theory for the Luttinger liquid\cite{OLL,PhysRevLett.92.237003,PhysRevB.72.073305,PhysRevLett.97.076401,Nattermann2008,DallaTorre2010,DallaTorre2012,PhysRevLett.113.260403}. 
The primary result will be the creation of a temporally non-local density-density interaction term, which is also quadratic in $\phi$. 
Hence, we are able to compute various two-point correlation functions of $\phi$ nonperturbatively. 
Comparing these correlation functions to those of the `closed' Luttinger liquid, we find that the dissipative bath destroys the perfect metallic conductivity of the Luttinger liquid, and also makes cosine terms far more relevant, in agreement with Ref. \citenum{PhysRevLett.97.076401}. 
The latter renders the system more sensitive to spatial potentials coupling to the density, as well as back-scattering processes, each of which further localizes the system. 
Although the CL bath thermalizes with the Luttinger liquid, in surprising contrast to Markovian baths, the decohering effect of the CL bath renders the Luttinger liquid more localized.

\section{Model} \label{sec:Model}
The `system' is a quantum fluid in a one dimensional ring of length $L$. 
For concreteness, we take the fluid's microscopic \dof~to be fermions, though our results should hold for bosons as well\cite{giamarchi_book}. 
In one dimension, the coarse grained theory of either is the Luttinger liquid\cite{Luttinger,MattisLieb,LutherPeschel,Haldane2,HaldaneLuttinger,giamarchi_book,vDS}, with Euclidean action
\be S^{\vphantom{\dagger}}_{LL} = \frac{\hbar u}{2 \pi K} \int\limits_{0}^{\beta \hbar} d\tau \, \int\limits_{-L/2}^{L/2} dx \,  \left( \left( \frac{1}{u} \partial_{\tau} \phi \right)^2 + \left( \partial_{x} \phi \right)^2 \right) \label{eq:LLactionxt} \,, ~ \ee 
where the velocity, $u$, and stiffness, $K$, are given by
\be u \equiv \left[ \left( v^{\vpd}_F + g^{\vps}_4 \right)^2 - g_2^2 \right]^{1/2} \, ,~ K \equiv \left[ \frac{v^{\vps}_F + g^{\vps}_4 - g^{\vps}_2}{v^{\vpd}_F + g^{\vps}_4 + g^{\vps}_2} \right]^{1/2} ~~, \label{eq:vgdef} \ee
where for spinless fermions, $g^{~}_2$ is the small-$q$ matrix element of the interaction between fermions near opposite Fermi points, and $g^{~}_4$ corresponds to fermions near the same Fermi point\cite{HaldaneLuttinger,giamarchi_book}. 
We have taken the usual limit of $\delta$-function interactions, but it is possible to treat any sufficiently short-ranged interaction\cite{HaldaneLuttinger}. 

The field $\phi$ is related to the fermion density via
\be \label{eq:rhoexact} \rho \left( x \right)  = -\frac{1}{ \pi} \nabla \phi \left( x \right) + \frac{1}{\pi \alpha} \cos \left[ 2 k^{\vpd}_F x - 2 \phi \left( x \right) \right] + \dots, ~~\ee
where we have omitted higher harmonics of $\phi$\footnote{The explicit form of these terms may depend on the bosonization convention and microscopic details, which are not of central importance in this work. 
This will also change for the chiral or helical Luttinger liquids.}, and $\alpha$ is a vanishing length scale (UV cutoff). 
The conjugate momentum to $\phi$ is $\Pi =\pi^{-1} \nabla \theta$, proportional to the \emph{current}. An equivalent action is given by replacing $\phi \leftrightarrow \theta$ and $K \leftrightarrow K^{-1}$ in \eqref{eq:LLactionxt}. 

In general, other terms in \eqref{eq:LLactionxt} are possible, e.g. if one considers curvature of the electron dispersion $\varepsilon\left( k \right)$, or various potentials that couple to the harmonic parts of the density \eqref{eq:rhoexact}. 
We neglect these at long wavelengths\cite{HaldaneLuttinger,giamarchi_book} to highlight the \emph{leading} alteration to the Luttinger liquid induced by dissipation. 
We consider the most relevant of these corrections using perturbative RG in Sec. \ref{sec:RGcos}.

The \emph{full} model consists of the action \eqref{eq:LLactionxt} for the `system', that of the oscillator bath, $S^{~}_B$, and a coupling, $S^{~}_C$, between the two:
\be S^{\vpd}_{\rm full} \left[ \phi, \varphi \right]= S^{\vpd}_{LL} \left[ \phi \right] + S^{\vpd}_B \left[ \varphi \right]+ S^{\vpd}_C \left[ \phi, \varphi \right], \label{eq:Sfull} \ee
where in the continuum one has\cite{CL1,CL2,CL4,LeggettTwoState,Guinea1}
\be S^{\vpd}_B \left[ \varphi \right] = \frac{1}{2} \int\limits_{0}^{\beta \hbar} d\tau \, \int\limits_{-L/2}^{L/2} dx \,  \left( \left( \frac{1}{c} \partial_{\tau} \varphi \right)^2 + \left( \partial_{x} \varphi \right)^2 \right) \label{eq:Sbath} \,, ~ \ee
for some velocity $c$. 
The system-bath coupling is linear in both the density, $\rho$ \eqref{eq:rhoexact} and the bath field, $\varphi$, 
\be S^{\vpd}_C \left[ \phi, \varphi \right] =  \int\limits_{0}^{\beta \hbar} d\tau \, \int\limits_{-L/2}^{L/2} dx \, \lambda \left( x \right) \rho \left( x \right) \partial_x \varphi ~.~~ \label{eq:Scouple} \ee
for arbitrary $\lambda$. 
Generally, our analysis will not be sensitive to microscopic details of either $S^{~}_B$ \eqref{eq:Sbath} or $S^{~}_C$ \eqref{eq:Scouple}, i.e. $c$ and $\lambda (x )$; however some details are essential for the elimination of the bath \dof~to be tractable. 
To wit, we require $S^{~}_B$ \eqref{eq:Sbath} to be quadratic, and that $S^{~}_C$ \eqref{eq:Scouple} be linear at least in the bath field $\varphi$. 
Both stipulations are inherent to the CL model; and most naturally captured by \eqref{eq:Sbath} and \eqref{eq:Scouple}.

Regarding the corresponding partition function $\mathcal{Z} = {\rm Tr} \left\{ e^{- S^{~}_{LL} - S^{~}_B - S^{~}_C} \right\}$, we can integrate (or trace) out the bath \dof, $\varphi$, to obtain a modified theory for the Luttinger liquid\cite{CL1,CL2,Guinea1,FeynmanVernon,CL4,LeggettTwoState,FZ}. 
This results in an ``effective'' term $S^{~}_{\rm eff} \left[ \phi \right]$ being added to $S^{~}_{LL}$ \eqref{eq:LLactionxt}, which will also be bilinear in $\phi$, to produce the Gaussian action
\be \label{eq:Seff} S^{\vpd}_{0} \left[ \phi \right] = S^{\vpd}_{LL} \left[ \phi \right]+ S^{\vpd}_{\rm eff} \left[ \phi \right] . \ee
This relies crucially on the linearity of \eqref{eq:Scouple} in both $\varphi$ and $\rho \sim - \frac{1}{\pi}\nabla \phi$, and $S^{~}_B$ being quadratic in $\varphi$; because this term results from Gaussian integration over $\varphi \left( x, \tau \right)$, it will always be \emph{non-local in time}\cite{OLL,PhysRevLett.92.237003,PhysRevB.72.073305,PhysRevLett.97.076401,Nattermann2008,DallaTorre2010,DallaTorre2012,PhysRevLett.113.260403,CL4,LeggettTwoState}. 

In simpler models---where the ``system'' consists of a single \dof, e.g. the position of a particle or a two-level system\cite{CL1,CL2,CL4,FZ,LeggettTwoState,Guinea1}---one makes a particular Ansatz for the bath spectral function. 
Because $S^{~}_B$ is Gaussian, this choice corresponds directly to a particular coupling $J \left( k, \omega \right)$ (in Fourier space) for the non-local term $S^{~}_{\rm eff}$ \eqref{eq:S1komega}. 
Absent such an Ansatz, $S^{~}_{0}$ would be sensitive to non-generic microscopic details of the bath; additionally, this choice affects only the $\omega$- and $k$-dependence of the \emph{coefficient} $J$ of the term(s) in $S^{~}_{\rm eff}$, whereas bilinearity in $\phi$ and temporal non-locality are jointly guaranteed by the linear and quadratic nature of \eqref{eq:Scouple} and $S^{~}_B$ in $\varphi$, respectively. 
Following this standard practice in the treatment of CL baths\cite{CL1,CL2,Guinea1,CL4,LeggettTwoState,FZ}, we assert \emph{some} spectral function (i.e. coupling $J \left( k, \omega \right)$ in $S^{~}_{\rm eff}$), and invoke RG relevance to constrain which forms merit consideration.

Thus, in general, one expects a term of the form \cite{OLL,PhysRevLett.97.076401}
\be \label{eq:S1komega} S^{\vpd}_{\rm eff} \left[ \phi \right]= J \left( k, \omega \right)  | \phi^{\vphantom{\dagger}}_{\omega,k} |^2 ~.~~\ee
Taylor expanding $J$ in $\omega$ and $k$ recovers a sum of terms $J_{m,n} \left| \omega \right|^m \left| k \right|^n$. 
Terms with $m+n > 2$ are \emph{irrelevant} in the RG sense, vanishing at long wavelengths. 
Terms with $m+n = 2$ are \emph{marginal}, i.e. fixed under the RG, and trivially modify the Luttinger liquid's velocity, $u$, and stiffness, $K$. 
Since CL baths generate interactions that are nonlocal in time, there should be at least one power of $\omega$; thus, we restrict to the ``Ohmic'' form, $J \propto \abs{\omega}$, to capture the relevant, non-trivial physics of the bath\footnote{One can regard the parameters $u$ and $K$ as having been modified to absorb any `marginal' terms}, as established in Ref. \citenum{PhysRevLett.97.076401}. 
We add to $S^{~}_{LL}$ \eqref{eq:LLactionxt} the term
\be \label{eq:OhmicEffAct} S^{\vpd}_{\rm eff} \left[ \phi \right] = \frac{\hbar \, \eta }{2 \pi} \int dk \int d\omega \abs{\omega} \left| \phi_{k,\omega} \right|^2, \ee
and restrict to $\eta \geq 0$\cite{CL1,CL2,FZ,QBM,OLL,Nattermann2008,DallaTorre2010,DallaTorre2012,PhysRevLett.97.076401}. 

A quadratic term of the form \eqref{eq:OhmicEffAct} will always be present in the presence of dissipation, however in general other terms in  $S^{~}_{\rm eff}$ may be realized upon integrating out the bath\cite{OLL,Nattermann2008,DallaTorre2010,DallaTorre2012,PhysRevLett.97.076401}. 
In particular, for spinless fermions, $\rho$ includes harmonics of $\phi$ \eqref{eq:rhoexact}; if the full form of $\rho$ is coupled to the bath, one expects a temporally non-local term $\propto \csc^2 \left( \tau - \tau' \right) \, \cos \left[ \phi \left( x, \tau \right) - \phi \left( x, \tau' \right) \right]$, and possibly higher harmonics\cite{Nattermann2008,PhysRevLett.97.076401}. 
However, the important physics of dissipation is already captured by the quadratic term \eqref{eq:OhmicEffAct}, and not only can we not treat the cosine term exactly, but its non-local nature complicates slightly the use of RG techniques, which are generally local. 

Thus, taking $\beta \hbar$ and $L$ finite for generality, the effective action for the Luttinger liquid with the bath \dof~integrated out takes the form
\be \label{eq:mainact0} S^{\vpd}_{0}  =  \sum\limits_{k,\omega^{\vps}_n} \frac{u^2 k^2 + \omega^2_n + u K J \left(k,\omega^{\vps}_n \right) }{2 \pi \, u \, K \, \beta \, L} \abs{\phi \left( k, \omega^{\vps}_n \right)}^2~,~~  \ee
as appears in Ref. \citenum{PhysRevLett.97.076401}. 
For analytic $J$, we invoke RG relevance and the arguments above to restrict our consideration to $J \left(k, \omega_n \right) = \eta \abs{\omega_n}$ \eqref{eq:OhmicEffAct} in the remainder.

\section{Two-point correlation functions} \label{sec:2ptcorr}
The Luttinger liquid's Gaussian action \eqref{eq:LLactionxt} allows for the computation of many physical properties directly from two point correlation functions, $\expval{ \phi \left( x, \tau \right) \phi \left( 0, 0 \right) }$. 
Thus, most of the calculational ``heavy lifting'' will be contained in this section, as we recover analytic solutions for the two point function without further approximation, even at non-zero temperature. 
We define the following:
\begin{align} F \left( x, \tau \right) &= K^{-1} \expval{ \left[ \phi \left( x, \tau \right) - \phi \left( 0, 0 \right) \right]^2 }  \label{eq:defF0}\\
&= \frac{2\pi u }{\beta \hbar L} \sum_{k} \sum_{\omega_n} \frac{ 1 - \cos \left( kx + \omega_n \tau \right)}{u^2 k^2 + \omega_n^2 + u K J \left( k, \omega_n \right)} \label{eq:defF1} \\ 
G \left( x, \tau \right) &= K^{-1} \expval{ \phi \left( x, \tau \right) \phi \left( 0, 0 \right) }  \label{eq:defG0}\\
&= \frac{ \pi u }{\beta \hbar L} \sum_{k} \sum_{\omega_n} \frac{  \cos \left( kx + \omega_n \tau \right)}{u^2 k^2 + \omega_n^2 + u K J \left( k, \omega_n \right)} \label{eq:defG1}, 
\end{align}
such that
\be \label{eq:FfromG} F \left( x , \tau \right) = 2 G \left( 0, 0 \right) - 2 G \left( x, \tau \right), \ee
where strictly speaking, $G \left( 0, 0 \right)$ is evaluated by first sending $x, u \tau \to \alpha$, and subsequently taking the limit $\alpha \rightarrow 0$ where safe\cite{giamarchi_book}. 
The parameter $\alpha$ is like a lattice spacing, and $\alpha^{-1} = \Lambda$ is a UV cutoff.

We calculate these correlation functions following the same procedure as for the Luttinger liquid without dissipation (i.e. $\eta=0$)\cite{giamarchi_book}. 
Surprisingly, we find exact solutions for both $G \left( x , \tau \right)$ and $F \left( x , \tau \right)$ for arbitrary $T$. 
Because the calculation is standard, and the results obtained are exact, we relegate the mathematical derivation to Appendix \ref{sec:mainGcalc}. 
We also confirm that taking $\eta \to 0$ reproduces the known results for the standard case in Appendix \ref{sec:matchclosed}. 
Finally, while we do not write them down explicitly, we note that \emph{single particle} correlation functions obtain directly from the results of this section, combined with those for the `closed' Luttinger liquid\cite[Appendix C]{giamarchi_book}. 
All of the results presented in this section correspond to the $L \to \infty$ limit, with $J \left( k , \omega \right) = \eta \abs{\omega}$.

\subsection{Zero temperature correlation functions} \label{subsec:T0corr}
At zero temperature, we send $\beta \to \infty$, simplifying the calculation in Appendix \ref{sec:mainGcalc}, as the sums over Matsubara frequencies in (\ref{eq:defF0}-\ref{eq:defG1}) become integrals. 
Regarding \eqref{eq:Gexact}, all of the terms containing factors $e^{- m \beta \dots}$ for $m \geq 1$ will vanish as $\beta \to \infty$, leaving only the $\beta$-independent terms. 
Thus, we have
\be \label{eq:GexT0} G \left( x, \tau \right)  =    \sum\limits_{n=0}^{\infty} \frac{e^{- \coup \, u \abs{\tau}  }}{2 \, n!} \left(\frac{-\coup \, x^2}{2 \, u \abs{\tau} } \right)^n K^{\vps}_{n} \left[ \coup \,  u \abs{\tau} \right], ~~~~~~~ \ee
where $K^{\vps}_n$ is a modified Bessel function of the second kind, and we use the shorthand
\be \label{eq:coupdef} \coup \equiv \eta \, K /2 \ee
throughout. 
Unlike the closed Luttinger liquid, the correlation function $G$ is not divergent in general. 
In most regimes, the sum over $n$ converges rapidly; in all cases, $G$ decays sharply to zero for $x, u \tau \gtrsim \coup^{-1}$.
 
The correlation function $F$ is given straightforwardly from the above using the relation \eqref{eq:FfromG} and the derivation of $G \left( 0, 0 \right)$ in Appendix \ref{sec:G00}. 
We have at zero temperature $F \left( x, \tau \right)=$
\be -\gamma - \ln \, \frac{\coup \, \alpha}{2} -   \sum\limits_{n=0}^{\infty} \frac{e^{- \coup \, u \abs{\tau} }}{n!} \left(\frac{-\coup x^2}{2  u \abs{\tau} } \right)^n K^{\vps}_{n} \left[ \coup  u \abs{\tau} \right]~,~~~~\label{eq:FexT0} \ee
where $\gamma$ is the Euler-Mascheroni constant, and implicit is the limit $\alpha \to 0$.

Although these results are exact---in the sense that no approximations were necessary beyond those outlined in our formulation of the model in Sec. \ref{sec:Model}---it is worth considering approximate forms of \eqref{eq:GexT0} corresponding to various physical regimes. 
A handful of limits can be taken straightforwardly; however, some must be analyzed with additional care. 
For example, the limits of $\eta \to 0$ and $\tau \to 0$ should not be taken independently of any others, and the limit $x \to \infty$ can only be taken along with some limit of $\eta$ or $\tau$.

\paragraph{Long time limit.---}Perhaps the most natural; in the limit of large argument, one has for the Bessel function in \eqref{eq:GexT0}
\be \label{eq:Klargez} \lim\limits_{z \to \infty} K^{\vps}_{\nu} \left( z \right) = \sqrt{\frac{\pi}{2z}} e^{-z} \, ,~~ \ee
and inserting this into the definition of $G \left( x, \tau \right)$ \eqref{eq:GexT0} gives
\be  G \left( x, \tau \right)  \approx  \sqrt{\frac{\pi}{8 \coup \, u \abs{\tau} }} \exp \left(- 2 \coup u \abs{\tau} - \frac{\coup \, x^2}{2 u \abs{\tau}} \right) \, ,~~~\label{eq:GT0longtime} \ee
where clearly the conformal invariance of the $\eta=0$ Luttinger liquid has been destroyed. 
The exponential decay in $t$ suppresses correlations for $\coup u \tau \gtrsim 1$; the remaining factors in \eqref{eq:GT0longtime} resemble a diffusion kernel, with diffusion constant $\propto u/\coup$. 
Thus, a non-zero density introduced at the origin, $x=0$, at time $\tau = 0$ will spread diffusively under the combined system and bath dynamics \eqref{eq:mainact0}, with exponential suppression on a time scale $1 / u \coup$.

\paragraph{Auto-correlation limit.---}Also of interest is the limit $x \to 0$, corresponding to a temporal auto-correlation function. 
Only the $n=0$ term in the sum in \eqref{eq:GexT0} survives, i.e.
\be \label{eq:GT0auto} G \left( 0, \tau \right) =  \frac{e^{- \coup \, u \abs{\tau}  }}{2} K^{\vps}_{0} \left[ \coup \,  u \abs{\tau} \right]~. ~~~\ee

For nonzero $x \ll 1$, we retain terms with $1<n<n_{*}$, resulting in an expansion to order $x^{2n_*}$; we then expand in the Bessel function's argument for further insight. 

The small argument limit is uninteresting: the result is  parametrically close in $x, \tau$ to $G (0,0)$ \eqref{eq:G00b}. At $\tau = 0$, we reinstate the $\alpha$-dependent convergence factor, per Appendix \ref{sec:G00}, and expand in $z = \coup \, u \abs{\tau}$
\be \label{eq:Besselsmall} K^{\vps}_0 \left( z \right) \to \sum\limits_{k=0}^{\infty} \frac{ \left(z/2\right)^{2k}}{\left( k!\right)^2 } \left( \psi \left( k+1 \right) - \ln \left(\frac{z}{2}\right) \right) ~, ~~~\ee
where $\psi$ is the DiGamma function.
Higher order corrections can be found following Appendices \ref{sec:G00} and \ref{sec:smalleta}.

The late time limit is unambiguous:
\be \lim\limits_{\tau \to \infty} G \left( 0, \tau \right) = 0~,~~ \label{eq:vacseize} \ee
referred to as ``seizing of the vacuum''\cite{OLL,KogutSusskind}. 
One can show that \eqref{eq:vacseize} holds for real time $t$ starting from \eqref{eq:defG1} with $\tau = it$, using contour integrals rotated $90^{\circ}$ in the complex plane, or by analytic continuation of \eqref{eq:GexT0} to real time. 
This seizing signals localization of the fermion \dof\cite{OLL,KogutSusskind}, and will be discussed in Sec. \ref{sec:seize}.

\paragraph{Various interaction limits.---}Details of the bare interactions in the Luttinger liquid are encoded in the velocity, $u$, and Luttinger parameter, $K$, via the parameters $g^{~}_{2,4}$ in \eqref{eq:vgdef}. 
The limit $K \to 1$ corresponds to free fermions; other limits, such as $K \to 0$ and $K \to \infty$, may be realized experimentally. 
Referring to the action \eqref{eq:LLactionxt}, the Luttinger liquid may be described in terms of \emph{either} the field $\phi$ or its dual, $\theta$; the corresponding bare actions \eqref{eq:LLactionxt} have overall coefficient $K^{-1}$ and $K$, respectively. 
The two-point functions of $\phi$ and $\theta$ correspond to the functions $G,F$, as the case may be, multiplied by $K$ and $K^{-1}$, respectively. 
In the extreme limits $K \to 0$ and $K^{-1} \to 0$, one of these will be zero and the other infinite: na\"ively, dissipation is unimportant in either limit; however, consideration of these scenarios likely requires a more careful treatment beyond the scope of this work.

Curiously, nothing in particular happens to \eqref{eq:GexT0} in the free fermion limit, $K =1$. 
Rather, it seems any interesting interaction effects must be encoded in $\cos \left[ \phi \right]$ terms of the type mentioned in Sec. \ref{sec:Model}, which we examine in Sec. \ref{sec:RGcos}. Otherwise, the non-interacting limit merely amounts to a specific value of $\coup = \eta / 2$ and $u = v^{\vps}_F$, the Fermi velocity, with nothing remarkable at the level of density-density correlations.

\paragraph{Weak coupling regime.---}Expansion of \eqref{eq:GexT0} to $O (\eta)$ is straightforward, as detailed in Appendix \ref{sec:smalleta}. 
Let us consider $F \left( x, \tau \right)$ to order $\eta^2$: at lowest order we recover the dissipationless result (see Appendix \ref{sec:matchclosed}),
\be \label{eq:Fclosedagain} F \left( x, \tau \right) = \frac{1}{2} \ln \left[ \frac{x^2 + u^2 \tau^2}{\alpha^2} \right]~,~~~\ee
then corrections from the factor $\exp \left( - \coup u \abs{\tau} \right)$, 
\be + \left( \frac{1}{2} \coup^2 u^2 \tau^2 - \coup u \abs{\tau} \right) \left( \gamma + \frac{1}{2} \ln \left[ r^2 \right] \right)~,~~~\label{eq:FetaExpTerms} \ee
and finally, from the Bessel function
\be \label{eq:FetaBesselTerms} + \frac{\coup^2}{4}\left( x^2 + u^2 \tau^2 \right) \left( \gamma - 1 + \frac{1}{2} \ln \, \frac{x^2 + u^2 \tau^2}{4 \alpha^2} \right) ~,~~~\ee
and we note that to order $\coup^2 \propto \eta^2$, only the terms arising from expansion of the overall factor $\exp \left( - \coup u \abs{\tau} \right)$ spoil the conformal invariance present for $\eta = 0$.

\subsection{Finite temperature correlation functions} \label{subsec:finTcorr}
As shown in Appendix \ref{sec:mainGcalc}, finite temperature correlations are no more difficult to obtain. 
For $T>0$,  $G\left( x, \tau \right) $ and $F\left( x, \tau \right) $ contain the respective terms \eqref{eq:GexT0} and \eqref{eq:FexT0} present at $T=0$, i.e.
\be G \left( x, \tau ; T \right) = G \left( x, \tau ; 0 \right) + \sum_{m=1}^{\infty} \sum_{\pm}  G \left( x, m \beta \hbar \pm \abs{\tau}\right)~,~~\label{eq:GfinTterms} \ee
with $G \left( x, \tau ; 0 \right)$ given by \eqref{eq:GexT0}. 
The dominant contribution at high temperature corresponds to $m=1$. 

By analogy, we obtain $F \left( x, \tau ; T \right)$ by adding to $F \left( x, \tau ; 0 \right)$ \eqref{eq:FexT0} the terms
\be \sum_{\substack{m=1, \\ \pm}}^{\infty} \left\{ \frac{e^{- \coup  u m \beta \hbar }}{2}  K^{\vps}_{0} \left[ \coup  u  m \beta \hbar  \right]  
-  2 G \left( x, m \beta \hbar \pm \abs{\tau}\right) \right\} ,~~~\label{eq:FfinTterms} \ee
which is difficult to parse, even restricted to $m=1$. 
Note that exponential factors in $G$ and $F$ that grow (rather than decay) in $\abs{\tau}$ are at most unity, since $0 \leq \tau < \beta \hbar$.

As for $T=0$, one can evaluate a number of physical limits for $T>0$; however, apart from the results of Sec. \ref{subsec:T0corr}, which are still present for $T>0$, little can be said about the finite temperature terms in the limits considered in Sec. \ref{subsec:T0corr} without expanding in $T$. 
Since the Luttinger liquid picture breaks down at high energies, only the low temperature limit is reasonable; that limit is well-captured by the results of Sec. \ref{subsec:T0corr}. 

Additionally, it is unclear how (or whether) to take the $\tau \to \infty$ limit of expressions involving $m \beta \hbar - \abs{\tau}$, since our recovery of Bessel functions $K_n \left[ z \right]$ is only valid if ${\rm Re} \left( z \right) >0$. 
Given that $0 \leq \tau \leq \beta \hbar$, this does not pose an issue for the result itself; however, for finite $\beta$, the ``long time'' limit is more subtle. 
For the purposes of transport and ``seizing'', we are interested in the limit of \emph{real time} $t \to \infty$, which we will address e.g. in Sec. \ref{sec:seize}.

\subsection{Vertex operator correlations} \label{subsec:expcorr}
In this section we consider two-point functions of ``vertex operators'', related, e.g., to the fermion creation/annihilation operators. 
These are exponential correlation functions of the form
\begin{align} C \left( x, \tau ; m \right) &=  \frac{1}{\left( 2 \pi \alpha \right)^2} \expval{ e^{i m \phi \left( x, \tau \right) }e^{-i m \phi \left( 0, 0 \right) } }  \label{eq:defC0}\\
&= \frac{1}{\left( 2 \pi \alpha \right)^2} \exp \left\{ - \frac{K \,m^2}{2} F \left( x, \tau \right) \right\} \label{eq:defC1} ~,~~\end{align}
where $F$ is given by \eqref{eq:FexT0} for $T=0$ and \eqref{eq:FfinTterms} for $T>0$. 

Unlike the closed Luttinger liquid ($\eta =0$), the absence of a constant, divergent term in $G \left( x, \tau \right)$ for $\eta >0$ allows for vertex operator correlations of the form
\bent\expval{ e^{i \sum_k A_j \, \phi \left( x_j, \tau_j \right) } } \eent
to be non-zero even for $\sum_k A_j \neq 0$. 
This will affect the use of standard techniques, e.g. Giamarchi-Schulz RG\cite{GiamarchiSchulz}, for $\cos \, \left[ m \phi \right]$ terms perturbing the action \eqref{eq:LLactionxt}. 

Let us compare $C \left( x, \tau \right)$ to its $\eta = 0$ form, $C^{\vps}_0 \left( x, \tau \right)$; taking $T=0$ and $m = 2$ for the first harmonic, to order $\eta^2$ (see Sec. \ref{subsec:T0corr} and App. \ref{sec:smalleta}) one has
\begin{gather} C^{\vps}_{\eta} \left( x, \tau \right) = C^{\vps}_0 \left(x,\tau \right) e^{K \coup^2 \left( x^2 + u^2 \tau^2 \right)/2} \notag \\
\times \left( \frac{4 e^{-2\gamma} }{ x^2 + u^2 \tau^2 }\right)^{\frac{K\coup^2}{4} \left(3 u^2 \tau^2 + x^2 - 4 u \abs{\tau}/\coup \right) } \label{eq:Csmalleta} ~,~~\end{gather}
where analysis of the behavior as $x,\tau \to \infty$ is complicated by competing terms. 

However, we can see the limiting behavior at large distances [times] directly from \eqref{eq:defC1}. For $C \left( x, \tau \right)$ of the form \eqref{eq:defC0}, note that $F \left( x, \tau \right) $  vanishes as \emph{either} $x,\tau \to \infty$ faster than $x^{-2}$ or $\left( u \tau \right)^{-2}$. In contrast to the dissipationless case, $C \left( x, \tau \right) \to 1$, rather than zero, for large $x$ or $\tau$. 

Because dissipation \eqref{eq:S1komega} destroys the conformal invariance of \eqref{eq:LLactionxt}, we cannot simply read off RG relevance of $\cos \, \left[ m \phi \right]$ terms perturbing \eqref{eq:LLactionxt} from the scaling dimension of their correlations, $C \left( x, \tau \right)$. Nevertheless, because $G \left(x , \tau \right)$ is finite as its arguments approach infinity, one expects that dissipation will generally render such cosine terms more relevant than for $\eta = 0$, since their correlations no longer vanish at long wavelengths.

\section{Seizing of the vacuum} \label{sec:seize}
An earlier prediction for Luttinger liquids coupled to a CL bath\cite{OLL} is a property termed \emph{seizing of the vacuum}\cite{KogutSusskind}, corresponding to localization of the bare fermion \dof~at $T=0$. Quantitatively, this is indicated by
\be \label{eq:seizedef} \lim\limits_{t \to \infty} G \left( 0, t \right) = 0~,~~\ee
for real time, $t$, and $G$ evaluated at $T=0$. This effect was reported in Ref. \citenum{OLL} for a similar model; the exact solutions of Sec \ref{sec:2ptcorr} confirm this property definitively.

It is easy to verify that the Euclidean time correlation function $ G \left( 0 , \tau \right) \to 0$ as $\tau \to \infty$ without caveat. Using analytic continuation, i.e. $\abs{\tau} = \tau \, {\rm sgn} \, \tau$, and thus $\abs{\tau}^2 = \tau^2 = \left( it \right)^2 = - t^2$, we note that $\abs{\tau}^{-n} K^{\vps}_n \left[ \coup u \abs{\tau} \right]$ has a series expansion involving only even powers of $\tau$. Using this, we can take the limit $t \to \infty$ safely, finding that the summand in \eqref{eq:GexT0} goes to zero, even without the help of the exponential decay $\exp \left( - \coup u \abs{\tau}\right)$. Additionally, it is possible to repeat the proceedings of Appendix \ref{sec:mainGcalc} for real time $t$, which requires the use of a rotated contour compared to Euclidean time derivation; nonetheless, taking $x \to 0$, one recovers an expression that unambiguously vanishes at large times, $t \to \infty$. 

Surprisingly, this behavior is not limited to the vacuum: the ``seizing'' effect, characterized by \eqref{eq:seizedef}, also holds for finite temperature $T>0$, and thus is present in \emph{excited states} as well. As for $T=0$, this can be seen either by analytic continuation of $G \left( x , \tau \right)$ to real time, or by reproducing the calculation of $G$ entirely for real time $t$. The latter requires taking $x=0$ at the outset, and taking $t \to \infty$ when safe. At finite $\beta$, the $t$-dependence dominates, and we see that $G \to 0$; as $\beta \to \infty$, one recovers the $T=0$ result, which also corresponds to seizing. Hence, we conclude that this effect is \emph{not} limited to the ground state, but is present throughout the spectrum.  However, at very high temperatures, one expects both a breakdown of the bosonization procedure itself, and for thermal fluctuations to outweigh this effect.

\section{Conductivity} \label{sec:Kubo}
We can also see evidence of localization from a transport calculation using the Kubo formula. Restricting to $1d$ fermions with electron charge $e$, the charge density is $\rho = -\frac{e}{\pi} \nabla \phi + \dots$; using the continuity equation, $\partial_t \rho + \nabla j = 0$, we have
\be \label{eq:currentoperator} j \left( x, t \right) = \frac{e}{\pi} \partial^{\vps}_t \phi \left( x, t \right)~,~~\ee
in the limit of vanishing source, from which we can compute the current in the presence of a weak source using linear response. The source will be a weak electric field oriented along the wire, given by $E \left( t \right) = E^{\vps}_0 e^{- i \left( \omega + i \delta \right) t }$, where $\delta$ is a small, positive real number inserted for convergence purposes. Writing $E = - \partial^{\vps}_t A$, where $A$ is the $1d$ vector potential, we invoke Ohm's law
\be\label{eq:ohmFT} j \left( q, \omega \right) = \sigma \left( q, \omega \right) E \left( q, \omega\right)~,~~\ee
where we will take $q\to 0$ to highlight the frequency dependence, and $j$ on the LHS is given by the expectation value of the current operator \eqref{eq:currentoperator}, which we compute using the Kubo formula,
\be  \expval{j \left( x, t \right)} =   \expval{j \left( x, t \right)}^{\vps}_0 +  \delta \expval{j \left( x, t \right)} \label{eq:kubo0}~.~~\ee

The first term in \eqref{eq:kubo0} is proportional to the source, $A^{\vps}_1$,
\be  \expval{j \left( x, \tau \right)}^{\vps}_0  = D A_1 \left( x, \tau\right) = -\frac{e^2 u K}{\pi \hbar}A_1 \left( x, \tau\right)~,~~\label{eq:diamagpart}\ee
known as the ``diamagnetic'' contribution. 
Going to Fourier space, the second term is given by
\be \label{eq:jfromchiFT}  \delta j \left( q, \omega \right) = \chi \left( q ; \omega \right) A_1 \left( q, \omega \right)~,~~\ee
and we can write Ohm's law \eqref{eq:ohmFT}
\be \sigma \left( q, \omega \right)  =  \frac{j \left( q, \omega \right) }{ E \left( q, \omega\right)} = \frac{D + \chi \left( q, \omega \right)}{i \left( \omega + i \delta \right)}~.~~\label{eq:ohmFT2} \ee
We next solve for $\chi$, which is a current-current correlation function:
\begin{align} \chi \left( q, \omega_n \right) &= -\frac{e^2 \omega^2_n}{\pi^2 \hbar}  \expval{  \phi^{*}_{q, \omega_n } \phi^{\vps}_{q, \omega_n } }  - D~,~~ \end{align}
and inserting this into \eqref{eq:ohmFT2}, and analytically continuing $i \omega_n = \omega + i \delta$ with $\delta= 0^{+}$ one finds
\begin{align} \sigma \left( q, \omega \right) &= \frac{e^2 }{\pi^2 i \hbar} \left( \omega + i \delta \right) \expval{  \phi^{*}_{q, \omega_n } \phi^{\vps}_{q, \omega_n } }  \label{eq:acconduct} \\ \sigma \left( q, \omega_n \right) &=  \frac{e^2 \omega_n}{\pi^2  \hbar} \frac{\pi u K}{u^2 q^2 + \omega_n^2 + \eta u K \abs{\omega_n}}~,~~\end{align}
and taking $q \to 0$, we recover the dc conductivity by further taking the limit $\omega \to 0$, i.e.
\begin{align} \sigma \left( \omega \to 0 \right)  &= \frac{e^2 u K}{\pi \hbar}\frac{1}{\delta - i \omega + \eta u K \, {\rm sgn} \left( \delta - i \omega\right)} \notag \\
&=  \frac{e^2}{\pi \hbar \eta \, {\rm sgn} \left( \delta \right) } = \frac{e^2}{\pi \hbar \eta}~,~~\label{eq:dcconductres} \end{align}
which is \emph{finite}. We also note a relation between \eqref{eq:dcconductres} and the condition for seizing, as discussed in Sec. \ref{sec:seize}. 

This result \eqref{eq:dcconductres} contrasts sharply with the infinite conductivity of the closed Luttinger liquid as $q, \omega \to 0$. This result does recover for $\eta \to 0$. Hence, we find further evidence that the dissipative bath localizes the underlying fermion excitations of the Luttinger liquid.

\begin{figure}[t!]
	\includegraphics[width = 0.95\columnwidth]{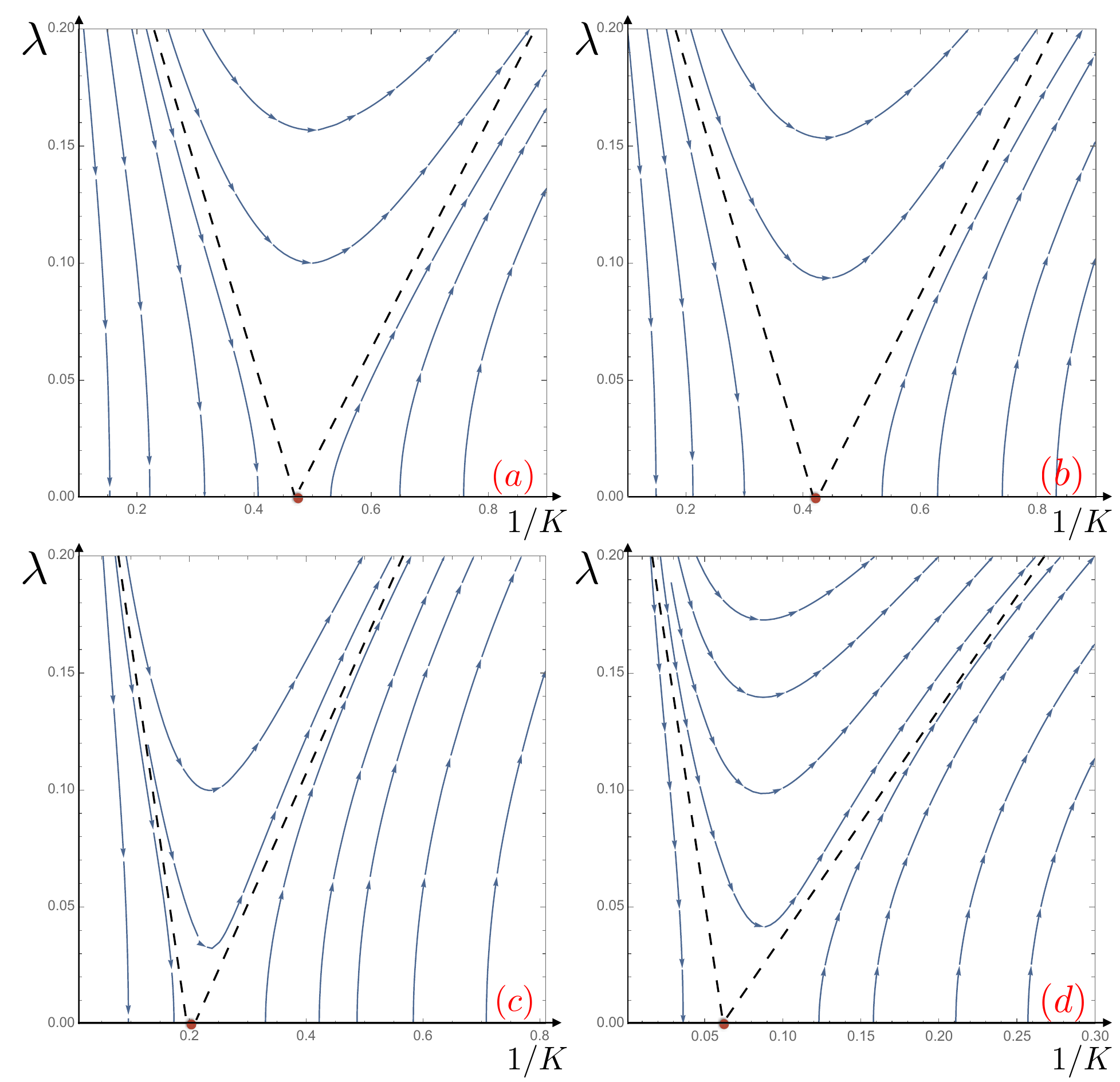} 
	\caption{\label{fig:KTRG} RG flow in the $\lambda$, $1/K$ plane, for various values of the dissipation strength, $\eta$;  $\lambda$ is the strength of the most relevant harmonic, corresponding to $m=2$ in \eqref{eq:harmpert}. We take the velocity $u$ and the coefficient $C$ in \eqref{eq:Kinvflow} to be unity; for $\eta = 0$, the Kosterlitz-Thouless transition corresponds to  $1/K^{*}$ = $1/2$. The panels $a$ through $d$ correspond respectively to $\alpha \eta = 10^{-3}$, $10^{-1}$, $1/2$, and $1$, where $\alpha = 1/\Lambda$ is a UV cutoff; note horizontal axis is different in each panel, the critical values $1/K^{*}$ in each are roughly $0.47$, $0.42$, $0.2$, and finally, $0.06$ for $\eta = \Lambda$. Under the RG, $\eta$ grows exponentially from its initial value, independent of $\lambda$ and $K$; for $\eta$ up to roughly one tenth the cutoff (panel $b$), the alteration to the dissipationless phase diagram is rather slight; for $\eta \gtrsim \Lambda/2$, we find a marked enhancement in the region of parameter space for which harmonic terms are relevant. }
\end{figure}

\section{Relevance of harmonic terms} \label{sec:RGcos}
We now consider the relevance of generic cosine terms, omitted from \eqref{eq:LLactionxt}, using standard momentum-shell RG. Our starting point is the action $S^{\vps}_0$, defined in \eqref{eq:Seff} and \eqref{eq:mainact0}, which contains the relevant, non-harmonic terms that one recovers after integrating out the bath \dof. We add to this a generic harmonic term of the form
\be S^{\vps}_1 \left[ \phi \right] = u \, \lambda^{\vps}_m \, \int\limits_{0}^{\beta \hbar} d \tau  \int\limits_{-L/2}^{L/2} dx \, \cos \left[ m \phi \left( x, \tau \right) \right] ~,~~\label{eq:harmpert} \ee
where $m \in \mathbb{N}$ specifies the harmonic: $m= 2$ is the first to appear in the density \eqref{eq:rhoexact}; an Umklapp (back-scattering) term corresponds to $m = 4$ (these will be multiples of two due to our convention). We defer analysis of multiple harmonics to future work.

We take $L, \beta \to \infty$ so the action \eqref{eq:mainact0} has an integral form in Fourier space as well,
\be \label{eq:mainact0int} S^{\vps}_0 = \int\limits_{-\Omega}^{\Omega} \frac{d\omega}{2\pi} \int\limits_{-\Lambda}^{\Lambda} \frac{dk}{2\pi} \frac{u^2 k^2 + \omega^2 + \eta K u \abs{\omega}}{2 \pi u K } \abs{\phi \left( k, \omega \right) }^2,~~~\ee
where $\Omega = u/\alpha$ and $\Lambda = 1/\alpha$ are frequency and momentum cutoffs, respectively.

The RG is implemented by separating the field $\phi \left( k, \omega \right) = \phi^{\vps}_s \left( k, \omega \right) + \phi^{\vps}_f \left( k, \omega \right)$, with `fast' modes living in an annulus in the momentum-frequency plane corresponding to $\left( b \alpha \right)^{-1} < q < \alpha^{-1}$, and all other modes `slow'. Note  $q = \sqrt{k^2 + \omega^2 / u^2}$, $\alpha$ is our usual short-distance cutoff, and $b = e^{\ell} \geq 1$ quantifies the extent of the coarse graining. For a given term, we trace out the `fast' modes, and then rescale frequency, momentum, and the fields themselves to obtain an effective theory of the slow \dof. The rescaling ($\tilde{\omega} = b \omega$, $\tilde{k} = b k$) is determined by the requirement that the `fixed' part of the action, $S^{\vps}_0$ remain unchanged under the RG, which requires that $\omega$ and $k$ have the same $b$, and results immediately in a rescaled coupling $\tilde{\eta} = b \eta$ and rescaled field $\tilde{\phi}$ given by $\tilde{\phi} \left( \tilde{k} , \tilde{\omega} \right) = b^{-2} \phi^{\vps}_s \left( k, \omega \right)$. In real space, one has $\tilde{x} = x/b$, $\tilde{\tau} = \tau/b$, and $ \tilde{\phi} \left( \tilde{x} , \tilde{\tau} \right) = \phi^{\vps}_s \left( x, \tau \right)$.

We now have the scaling of $\eta$; the RG for all other terms follows from the usual, $\eta=0$ case. We perform a cumulant expansion of $S^{\vps}_1$ to order $\lambda^2_{m}$, which for $\eta = 0$, gives rise to standard Kosterlitz-Thouless RG flow\cite{giamarchi_book}. All other differences compared to the closed case arise from modification of the `fast' two point function, taken with respect to $ S^{\vps}_0$, which includes also the quadratic dissipation term. That function, $K  G^{\vps}_{0,f} = \expval{ \phi^{2}_f \left( x, \tau \right) }^{\vps}_{0,f}$ is given by
\be G^{\vps}_{0,f} \left( x, \tau \right) =  \iint_{\abs{\omega}, \abs{k} \in f} \frac{d\omega}{2\pi} \, \frac{dk}{2\pi} \, \frac{\pi u \cos \left( kx + \omega \tau\right)}{u^2 k^2 + \omega^2 + u \eta K \abs{\omega} }~,~~\ee
which taking $x = r \cos \theta$, $u \tau = r \sin \theta$, $k = q \cos \psi$, $\omega = u q \sin \psi$, becomes
\be  G^{\vps}_{0,f} \left( x, \tau \right)= \frac{1}{4\pi} \int\limits_{\Lambda/b}^{\Lambda} dq \int\limits_{0}^{\pi} d\psi \frac{\cos \left( q r \cos \left( \theta - \psi \right) \right)}{q + \eta K \abs{\sin \psi }} ~,~~\ee
and details of the RG ensure that we need only consider $d G / d \ell$ as $\ell \to 0$ ($b \to 1$), i.e.
\be \frac{d G^{\vps}_{0,f}}{d\ell} = \frac{1}{4 \pi} \int\limits_0^{2\pi} d\psi \frac{ \cos \left( \frac{r}{\alpha} \cos \left( \theta - \psi\right) \right)}{1+ \alpha \eta K \abs{\sin \psi}}~,~~\label{eq:Gfderiv} \ee
evaluated either at $r=0$, or integrated directly over $\theta$, which we can therefore shift by $\psi$. In either case, the numerator's dependence on $\psi$ is eliminated, and we find a simple relation to the dissipationless result:
\be \frac{d G^{\vps}_{0,f}}{d\ell} \left( x, \tau, \eta \right) =  \nu \left( \alpha \eta K \right) \, \frac{d G^{\vps}_{0,f}}{d\ell} \left( x, \tau, \eta= 0  \right) ~,~ ~\label{eq:GGrelation} \ee
where 
\begin{align}  \nu \left(z \right) &= \frac{1}{2 \pi}  \int\limits_0^{2\pi} d\psi \frac{1}{1+ z \abs{\sin \psi}} \label{eq:nufunc0} \\
&= \frac{1}{\sqrt{1-z^2}} \left(1 - \frac{2}{\pi} {\rm tan}^{-1} \left[ \frac{z}{\sqrt{1-z^2}} \right] \right) \label{eq:nufunc} ~,~~\end{align}
which has important limits $\nu \left( 0 \right) = 1$, $\nu \left( 1 \right) = 2/\pi$, and $\nu \left( z \right) \to 0$ as $z \to \infty$.

Thus, the various couplings of this theory flow according to
\begin{align} \frac{d \eta}{d\ell} &= \eta \\
\label{eq:Uflow} \frac{d \lambda^{\vps}_m}{d\ell} &= \left( 2 - \frac{m^2 K }{4} \nu \left( \alpha \eta K\right) \, \right) \lambda^{\vps}_m \\
\label{eq:Kinvflow} \frac{d K^{-1}  }{d\ell} &= \pi \nu \left( \alpha \eta K \right) C^{\vps}_0 \left( K \right) m^2 \lambda^2_m~,~~\end{align}
where $C^{\vps}_0 \left( K \right)$ is strictly positive function: it can be recovered from the $\eta=0$ case, but its precise form is unimportant.

Compared to the usual RG flow with $\eta = 0$, here, we see that dissipation is strictly relevant, with $\eta$ diverging exponentially as the RG is run. The other two couplings, the stiffness (Luttinger parameter) $K$ and strength of the harmonic perturbation $\lambda^{\vps}_m$ initially follow the usual Kosterlitz-Thouless flow until $\eta$ approaches the cutoff, $\Lambda = 1/\alpha$, as can be seen in Fig.~\ref{fig:KTRG}. For $\eta \gtrsim \Lambda/10$, the harmonic terms become relevant for increasing ranges of $K$. Throughout, the overall strength of the Luttinger liquid action \eqref{eq:LLactionxt}, $K^{-1}$ also grows, with the rate of growth slowed as $\eta$ increases. 

Referring to \eqref{eq:nufunc}, for $\alpha \eta K = 1$, $\nu$ will only have decreased in value from unity to $2/\pi$ compared to the $\eta=0$ value; this gives the impression that any interesting effect due to dissipation only occurs asymptotically late in the RG. However, due to the dependence of $\nu$ on $K$ as well, dramatic changes occur when $\alpha \eta$ approaches unity, as depicted in the bottom right of Fig.~\ref{fig:KTRG}, whereupon the critical value of $K^{-1}$ for $\lambda^{\vps}_m$ is suppressed tenfold compared to $\eta=0$. If $\eta$ is allowed to grow beyond the cutoff, only extremely large values of $K$ can prevent harmonic terms from being relevant; for $\eta \gg 1/\alpha$, $K$ ceases to flow entirely, and \eqref{eq:Uflow} becomes
\be \label{eq:Uflowlate} \frac{d \lambda^{\vps}_m}{d\ell} \to 2  \lambda^{\vps}_m ~,~~\ee
such that \emph{all} harmonic terms are relevant. 

\section{Conclusion}
In summary, we have investigated the effect of a dissipative bath on the properties of the Luttinger liquid. We argued that the effective contribution to the Luttinger liquid action corresponding to Ohmic dissipation captures the generic, relevant physics of CL baths. For this action, we computed two-point correlation functions non-perturbatively and for arbitrary temperature, finding evidence that dissipation makes the system vulnerable to localizing potentials. We later confirmed this using an RG analysis of generic harmonic terms, which we find to be more relevant under coarse-graining in the presence of dissipation, if not always relevant. For a wide range of parameters, the physics of the Luttinger liquid is altered substantially. Finally, a simple transport calculation reveals that dissipation destroys the perfect conductivity of the Luttinger liquid even in the absence of any spatial potentials or disorder, via Zeno localization. All of these findings show that the bath effectively \emph{enhances} localization in this interacting quantum system, in contrast to the usual intuition from Markovian baths that decoherence ought to make a system less localized.

These surprising results invite follow-up study, which should be facilitated by the exact correlation functions presented here. Given the increased importance of harmonic terms in the presence of dissipation, one direction for follow-up work is to examine the temporally-nonlocal harmonic terms generated by integrating out the bath. Also, consideration of higher-body correlations, specialization to physical applications, and higher dimensions may be of interest.

\noindent \emph{Acknowledgements.---}~We thank Y.-Z. Chou and R. Vasseur in particular for substantial feedback during the completion of this work, as well as
R. Nandkishore, S. A. Parameswaran, and B. A. Ware for illuminating discussions. This research was supported by the National Science Foundation via Grants DGE-1321846 (Graduate Research Fellowship Program) and DMR-1455366. We thank the University of Colorado, Boulder and the Kavli Institute for Theoretical Physics at the University of California, Santa Barbara, which is supported from NSF Grant PHY-1748958, for their hospitality while parts of this work were completed.

\appendix

\section{Derivation of correlation function} \label{sec:mainGcalc}
This section details the calculation of the correlation function $ G \left( x, \tau \right)$ for the Luttinger liquid with dissipation, for any temperature. Starting from \eqref{eq:defG0}, we take $L\to \infty$ by necessity, and restricting to the \emph{Ohmic} case $J \left( k , \omega \right) = \eta \abs{\omega}$, we have
\be G \left(x, \tau \right) =  \frac{i u}{4\pi} \int\limits_{-\infty}^{\infty} dk 
\oint dz \,  \frac{\, h^{\vpd}_B \left(z \right)  \, \cos \left( kx -i \tau z \right)}{z^2 - u^2 k^2 - u K \eta \left| z \right|},~~~~~~~ \label{eq:ohmint} \ee
where $z = i \omega $ as usual, and using the shorthand $\coup\equiv K \eta/2$ \eqref{eq:coupdef}, the denominator of \eqref{eq:ohmint} has zeros at $z^{*}_{\pm}=\pm u Z^{\vphantom{\dagger}}_{k}$, with 
\be Z^{\vphantom{\dagger}}_{k} &= \coup + \sqrt{\coup^2 + k^2}~, \label{eq:evenpole} \ee
which follows straightforwardly from setting $z = R e^{i \psi}$ and solving $z^2 = u^2 k^2 + u K \eta \abs{z}$, where the RHS of that expression is real, constraining $\psi = n \pi / 2$. 

Contour integration over $z$  returns the residues from poles of $h^{\vpd}_B$, reproducing the sum over discrete Matsubara frequencies, as well as residues corresponding to the poles at $z^{*}_{\pm}$. This quantity vanishes when taken along the great circle $\left|z\right| \rightarrow \infty$, and thus the Matsubara sum is equal to minus the contribution from the $z^{*}_{\pm}$ poles, as for the `closed' Luttinger liquid. However, in the `open' case, the poles of the denominator of \eqref{eq:ohmint} do not simply give $Z_k^{-1}$, as was the case for $\eta = 0$ whence $Z_k = k$. The result is 
\begin{gather} G \left( x, \tau \right) =   \int\limits_0^{\infty} dk \, \frac{ \cos \left( kx \right)}{\sqrt{\tilde{\eta}^2 +k^2} } ~\times \notag \\
\left\{  n^{\vphantom{\dagger}}_B \left( u Z_{k}  \right)  \cosh \left( u \tau Z_{k}  \right)  + \frac{1}{2} e^{- u \left| \tau \right| Z_{k} } \right\}  \label{eq:OhmG1} \end{gather}
where $Z_k$ is strictly positive, and reduces to $k$ as $\eta \to 0$. We can massage the term in braces, noting that $\cosh \left( u \tau Z_{k}  \right) = \cosh \left( u \abs{\tau} Z_{k}  \right) = \frac{1}{2} \sum_{\pm} e^{\pm u \abs{\tau} Z_k}$, and that 
\be \label{eq:nBexp} n^{\vphantom{\dagger}}_B \left( u Z_{k}  \right)  = \frac{1}{e^{\beta \hbar u Z_k} -1} = \sum\limits_{m=1}^{\infty} e^{-m \beta \hbar u Z_k} \ee
meaning all of the terms in curly braces in \eqref{eq:OhmG1} can be written in the form $e^{-A \cdot u Z_k} $. Explicitly, the braced term is
\be \label{eq:OhmHelp1} \frac{1}{2} \left\{ e^{- u \left| \tau \right| Z_{k} } + \sum_{\pm} \sum\limits_{m=1}^{\infty} e^{- u \, \left( m \beta \hbar \pm \abs{\tau} \right)  Z_{k} } \right\} \, ,~~\ee
which simplifies the integration procedure substantially, as all of these terms have the same general form. For $T=0$, i.e. $\beta \to \infty$, the latter term is simply zero. We rewrite \eqref{eq:OhmG1} in the generic form
\be \label{eq:OhmG2} G \left( x, \tau \right) =  \frac{1}{2} \sum_{\sigma} \int\limits_0^{\infty} dk \, \frac{ \cos \left( kx \right)}{\sqrt{\tilde{\eta}^2 +k^2} } e^{- A^{\left( \tau \right)}_{\sigma}  Z_k} \, , ~\ee
where the $\tau$-dependent coefficients $A^{\vps}_{\sigma}$ reproduce the terms in \eqref{eq:OhmHelp1}, indexed by $\sigma$.

We next invoke hyperbolic substitution, $k \equiv \coup \sinh \left( \lambda \right) $, and therefore $dk \, = \coup \cosh \left( \lambda \right) d\lambda$ and $Z_k \rightarrow \coup \left( 1 + \cosh \left( \lambda \right) \right)$, and the integral in \eqref{eq:OhmG2} becomes
\be \int\limits_0^{\infty} d\lambda \, \cos \left(  \coup  x  \sinh \left( \lambda \right) \right) e^{-\coup  A^{\left( \tau \right)}_{\sigma} \left( 1+  \cosh \left( \lambda \right) \right)} \, , ~~ \label{eq:OhmG3} \ee
and we then invoke a Taylor expansion for the cosine, i.e.
\begin{gather} G\left( x, \tau \right) = \frac{1}{2}  \sum_{\sigma} e^{- \coup A^{\left( \tau \right)}_{\sigma}  }  \sum_{n=0}^{\infty} \frac{ \left( -1 \right)^n \left(\coup  x\right)^{2n}}{\left(2n\right)!} ~ \times \notag \\
\int\limits_0^{\infty} d\lambda \, \sinh^{2n} \left( \lambda \right) e^{-\coup  A^{\left( \tau \right)}_{\sigma}   \cosh \left( \lambda \right) } \, . ~~ \label{eq:OhmG4} \end{gather}

At this point, we make use of a particular integral representation of the modified Bessel function of the second kind,
\be K^{\vphantom{n}}_{n} \left( z \right) = \frac{\pi^{1/2} \left(z / 2 \right)^n}{\Gamma \left( n + \frac{1}{2} \right)} \int\limits_0^{\infty} dt ~  \left[ \sinh \left( t \right) \right]^{2n} e^{- z \cosh \left( t \right)} \, ,~~~~~ \label{eq:NIST1} \ee
or written more usefully,
\be \frac{\left(2 n \right) ! }{\left( 2 z \right)^n n!} K^{\vphantom{n}}_{n} \left( z \right) = \int\limits_0^{\infty} dt ~  \left[ \sinh \left( t \right) \right]^{2n} e^{- z \cosh \left( t \right)} \, , ~~ \label{eq:NIST2}  \ee
which we can use to express \eqref{eq:OhmG4} exactly as
\be G \left( x, \tau \right)  =  \sum\limits_{\sigma}  \sum\limits_{n=0}^{\infty} \frac{e^{- \coup A^{\left( \tau \right)}_{\sigma}  }}{2 \, n!} \left(\frac{-\coup \, x^2}{2 \, A^{\left( \tau \right)}_{\sigma}  } \right)^n K^{\vphantom{n}}_{n} \left[ \coup  \, A^{\left( \tau \right)}_{\sigma}\right], ~~~~~~~ \label{eq:Gexact} \ee
where $A^{\left( \tau \right)}_{\sigma}$ are summed over $A \left( \tau \right) = \alpha + u \abs{\tau}$ and $ A_{m,\pm} \left( \tau \right) =  u\,   \, \left( m \beta \hbar \pm \abs{\tau} \right) $ for  positive integers $m \geq 1$, and we have reinstated $\alpha$ as it would appear had we included the usual convergence factor $e^{-i \omega \alpha / u}$ starting from \eqref{eq:ohmint}. Strictly, this convergence factor ought to be included, as is standard practice even in the dissipationless limit. Unlike the $\eta=0$ case, for $\eta > 0$ all integrals converge, giving exact results, except at $\tau=0$, necessitating the convergence factor in the $\omega$ integral. Finally, for $T = 0$, only the former, $\beta$-independent term appears in \eqref{eq:Gexact}. 

\section{Evaluating $G \left( 0, 0 \right)$} \label{sec:G00}
The evaluation of the more generic correlation function $F \left( x , \tau \right)$ requires knowledge of $G \left( 0, 0 \right)$, which requires the $\alpha$-dependent versions of $A^{\left( \tau \right)}_{\sigma}$ in the previous section. We will take $\alpha \to 0$ wherever safe. Regarding \eqref{eq:Gexact}, note the limit $x \to 0$ can be taken safely, and only the $n=0$ term remains:
\be \label{eq:G00a}  G \left( 0,0 \right) =  \sum\limits_{\sigma}  \frac{e^{- \coup A^{\left( 0 \right)}_{\sigma}  }}{2}  K^{\vps}_{0} \left[ \coup  \, A^{\left( 0 \right)}_{\sigma}\right], ~~~~~~~ \ee
where now the sum over $\sigma$ of the terms $A^{\left( 0 \right)}_{\sigma}$ corresponds to the terms $A = \alpha$ and $A^{\vps}_{m, \pm } = \alpha + u m \beta \hbar \to  u m \beta \hbar $, i.e.
\be G \left( 0, 0 \right) =\frac{1}{2} K^{\vps}_0 \left[ \coup \alpha \right] + \sum\limits_{m=1}^{\infty} e^{- \coup u m \beta \hbar} K^{\vps}_0 \left( \coup u m \beta \hbar \right) , ~~~~~ \label{eq:G00b}  \ee
where only the first term survives at $T=0$, and we have already taken the $\alpha \to 0$ limit where safe. Referring to the exact series expansion for $K^{\vps}_0 \left( z \right)$, we note that the limit $z =\coup \alpha \to 0$ can be taken safely in the majority of terms, resulting in 
\be \label{eq:K0exp0} \lim\limits_{z \to 0} K^{\vps}_0 \left( z \right) = -\gamma-\lim\limits_{z \to 0} \ln \left(\frac{z}{2} \right) ~,~~~~ \ee
where $\gamma \approx 0.577216$ is the Euler-Mascheroni constant. 
\\

\section{Matching the closed case for $\eta \to 0$} \label{sec:matchclosed}
Also note that inserting the form of the expansion of $K^{\vps}_n$ for arbitrary index $n$ about zero argument into \eqref{eq:Gexact}, and taking the limit $\coup \propto \eta \to 0$ of $\coup^n K^{\vps}_n$ recovers exactly the results for the closed case\cite{giamarchi_book}. We recover from \eqref{eq:Gexact} in the limit $\coup \to 0$
\begin{gather} \lim\limits_{\coup \to 0 } G \left( x, \tau \right) =  \notag \\
- \frac{1}{2} \sum_{\sigma} \left\{ \gamma +  \ln \left( \frac{\coup}{2} \right) + \frac{1}{2} \ln \left( A^2_{\sigma} \left( \tau \right) + x^2 \right) \right\} \label{eq:Gclosedlim}  ~,~~~\end{gather}
with $A^{\left( \tau \right)}_{\sigma}$ defined as before. 

For $T=0$, we have only one allowed configuration $\sigma$ that corresponds to a non-vanishing term, with $A^{\left( \tau \right)}_{\sigma} \to \alpha + u \, \abs{\tau}$. Thus, in this limit \eqref{eq:Gclosedlim} becomes
\begin{gather} \lim\limits_{\coup,T \to 0 } G \left( x, \tau \right) = - \frac{1}{2} \lim\limits_{\coup \to 0 } \ln \left( \frac{\coup}{2} \right) - \notag \\
\frac{\gamma}{2} - \frac{1}{4} \ln \left[ \left( \alpha + u \, \abs{\tau} \right)^2+ x^2 \right] ~,~~~ \label{eq:Gallzerolims} \end{gather}
which resembles the result i for the closed case\cite{giamarchi_book}, though matching divergent constants is murky at best. However, we note that 
\be \label{eq:G00allzeros}  \lim\limits_{\coup,T \to 0 } G \left(0, 0 \right) = -  \frac{\gamma}{2} -\frac{1}{2} \lim\limits_{\coup \to 0 } \ln \left( \frac{\coup \, \alpha}{2} \right)~,~~\ee
and using now the formula for $F$ \eqref{eq:FfromG} in combination with \eqref{eq:Gclosedlim}, we have for $T=0$:
\be \label{eq:Fallzeros} \lim\limits_{\coup,T \to 0 } F \left(x,\tau \right) = \frac{1}{2} \ln \left[ \frac{u^2 \abs{\tau}^2 + x^2}{\alpha^2} \right]~,~~\ee
in perfect agreement with the standard result\cite{giamarchi_book}. At finite temperature, this procedure is more cumbersome, and we content ourselves with the benchmark \eqref{eq:Fallzeros} as ample validation of our results for $\eta \neq 0$.

\section{Expansion for small $\eta$} \label{sec:smalleta}
Moving slightly beyond $\eta = 0$ may provide some insight. We restrict here to $T=0$; while one can certainly repeat this procedure at finite temperature, it affords little insight beyond the exact results. Looking at the definition of $G \left( x, \tau \right)$ \eqref{eq:GexT0}, 
\bent G \left( x, \tau \right)  =    \sum\limits_{n=0}^{\infty} \frac{e^{- \coup \, u \abs{\tau}  }}{2 \, n!} \left(\frac{-\coup \, x^2}{2 \, u \abs{\tau} } \right)^n K^{\vps}_{n} \left[ \coup \,  u \abs{\tau} \right], ~~~~~~~ \eent
we can re-write this as
\be \label{eq:GsmallEtaNice} \frac{e^{-z}}{2} \sum\limits_{n=0}^{\infty} \frac{\left(-1\right)^n}{2^n \, n!} \left(\frac{x}{ u\, \tau } \right)^{2n} \left[ z^n \, K^{\vps}_{n} \left( z \right) \right], ~~~~~~\ee
where $z = \coup \, u \abs{\tau}$ is a useful shorthand, as we will take the $z \to 0$ limit (recall $\coup \propto \eta$ \eqref{eq:coupdef}). Note that the limit $\eta \to 0$ is unimportant to the evaluation of $G \left( 0, 0 \right)$ in Appendix \ref{sec:G00}. 

Regarding \eqref{eq:GsmallEtaNice}, we now evaluate the summand to order $z^2$ as $z \to 0$ (ignoring for now the term $e^{-z}$), noting that the Modified Bessel function $K^{\vps}_n \left( x \right)$ has a well known Maclaurin series in $z$. We have at zeroth order the terms recovered Appendix \ref{sec:matchclosed}, 
\be \label{eq:Geta0terms} \frac{e^{-\coup u \abs{\tau}}}{2} \left( - \gamma - \ln \left[ \frac{\coup}{2} \right] - \frac{1}{2} \ln \left[ x^2 + u^2 \tau^2 \right] \right)~,~~~\ee
the latter obtaining from contributions for terms at all $n$. The lowest terms arising from the summand at non-trivial order are proportional to $\coup^2$, i.e.
\be \label{eq:Geta2terms} \frac{\coup^2}{8} \left( x^2 + u^2 \tau^2 \right) \left\{ 1 - \gamma + \ln 2 -  \frac{1}{2} \ln \left[ x^2 + u^2 \tau^2 \right] \right\} ~,~~~\ee
where we have dropped the overall exponential term above, and $\gamma$ is the Euler-Mascheroni constant in both \eqref{eq:Geta0terms} and \eqref{eq:Geta2terms}. Notably, at low order only the overall factor of $e^{- \coup u \abs{\tau} }$ spoils the conformal invariance present without dissipation (the invariance is not present in each term in the summand of \eqref{eq:GsmallEtaNice} individually, but is restored by the various contributions from different terms at a given order in $\eta$). 

We have then for the correlation function $F \left(x , \tau \right)$ the following, expanding now the exponential decay term in $\eta$ as well:
\begin{gather} F \left( x , \tau \right)  = \frac{1}{2} \ln \left[ \frac{x^2 + u^2 \tau^2 }{\alpha^2} \right] \notag \\
+ \left( \frac{\coup^2 u^2 \tau^2}{2} - \coup u \abs{\tau} \right) \left( \gamma + \ln \left[ \frac{\coup}{2} \right] + \frac{1}{2} \ln \left[x^2 + u^2 \tau^2 \right]  \right) \notag \\
+ \frac{\coup^2}{4} \left( x^2 + u^2 \tau^2 \right) \left( \gamma - 1 - \ln 2 + \frac{1}{2} \ln \left[x^2 + u^2 \tau^2 \right]   \right) \label{eq:FsmallEta} ~,~~~\end{gather}
where the term on the RHS of the first line corresponds to $\eta = 0$, and we note that $\eta^z \ln \eta \to 0$ as $\eta \to 0$ for positive $z$, which eliminates the $\ln \coup$ term in the second line.

\bibliography{DLL.bib}

\end{document}